\documentclass[12pt]{article}
\begin{document}
\baselineskip 18pt
\newcommand{\Tr}{\mbox{Tr\,}}
\newcommand{\beq}{\begin{equation}}
\newcommand{\eeq}[1]{\label{#1}\end{equation}}
\newcommand{\bea}{\begin{eqnarray}}
\newcommand{\eea}[1]{\label{#1}\end{eqnarray}}
\renewcommand{\Re}{\mathrm{Re}\,}
\renewcommand{\Im}{\mathrm{Im}\,}
\begin{titlepage}
\hfill  CERN-TH/99-49 NYU-TH/99/2/04 hep-th/9903085
\begin{center}
\hfill
\vskip .4in
{\large\bf RG Fixed Points in Supergravity Duals of 4-d Field Theory and
Asymptotically AdS Spaces}
\end{center}
\vskip .4in
\begin{center}
{\large M. Porrati and A. Starinets\footnotemark}
\footnotetext{e-mail: massimo.porrati@nyu.edu,
andrei.starinets@physics.nyu.edu}
\vskip .1in
{\em Theory Division CERN, Ch 1211 Geneva 23, Switzerland}
\vskip .1in
and
\vskip .1in
{\em Department of Physics, NYU, 4 Washington Pl.,
New York, NY 10003, USA}
\end{center}
\vskip .4in
\begin{center} {\bf ABSTRACT} \end{center}
\begin{quotation}
\noindent
Recently, it has been conjectured that supergravity solutions with two
asymptotically $AdS_5$ regions describe the RG flow of a 4-d field theory
from a UV fixed point  to an interacting IR fixed point.
In this paper we lend support to this
conjecture by showing that, in the UV (IR) limit,
the two-point function of a minimally coupled scalar field depends only on
the UV (IR) region of the metric, asymptotic to $AdS_5$.
This result is consistent with the interpretation of the radial
coordinate of Anti de Sitter space as an energy scale, and it may provide an
analog of the  Callan-Symanzik equation for supergravity duals
of strongly coupled field theories.
\end{quotation}
\vfill
CERN-TH/99-49 \\
February 1999
\end{titlepage}
\eject
\noindent
The duality between gauge theories and (super)string geometries, first proposed
for conformal field theories~\cite{malda,gkp,w1}, also holds in a more general
setting. Particularly interesting is the case when the superstring geometry
is only asymptotically $AdS_5$. This setting describes 4-d gauge theories
that are (super)conformal only in the ultraviolet. In the infrared, they may
confine and/or screen charges~\cite{ks,m,g},
or reduce to another conformal field
theory~\cite{kw,gppz,dz}. The 5-d metric that describes the latter case has
two regions, respectively ``far'' and ``close'' to the brane,
where the metric is asymptotically $AdS_5$. The interpolating metric is  still
invariant under the 4-d Poincar\'e group and it reads~\cite{gppz}:
\beq
ds^2= e^{2\phi(z)}(dz^2 + \eta_{\mu\nu}dx^\mu dx^\nu),  \;\;\; \eta_{\mu\nu}=
diag  (-1,1,1,1) .
\eeq{m1}
For small $z$ (``far'' from the brane) the prefactor in the metric has the
following expansion:
\beq
\phi(z)= -\log \left( {z\over R_{{{UV}}}}\right) + O(z).
\eeq{m2}
Here, $R_{UV}$ is the radius of the ``far'' AdS region.
For large $z$ the expansion is, instead:
\beq
\phi(z)= -\log \left( {z\over R_{{{IR}}}}\right) + O(1/z).
\eeq{m3}
In ref.~\cite{gppz} it was shown that an interpolating metric as in
Eq.~(\ref{m1}) exists, and that it describes mass deformations of
N=4 $SU(N)$ super-Yang Mills
theory. The metric was shown to exist using some general properties of type
IIB, 5-d, gauged supergravity~\cite{grw1,grw2}. The proof given in
ref.~\cite{gppz} does not rely heavily on specific properties of gauged
supergravity, and it is valid also in a more general context;
in type 0 strings, for instance.

The interpolating metric was interpreted as describing the
renormalization group flow from an UV N=4 superconformal theory to an IR
conformal theory. This interpretation is suggested by the UV/IR
connection~\cite{m,sw,pp}, i.e. by the identification of the $AdS_5$
coordinate,$z$, with an appropriate length scale in the 4-d field theory.

A problem with a direct interpretation of the equations of motion of gauged
supergravity as RG equations is that they are second order, instead of first
order. This seems to suggest that $z$ cannot be identified with the
renormalization scale of the 4-d field theory.
Purpose of this paper is to prove that this identification
is nevertheless correct, namely, that the IR physics of the boundary field
theory is essentially independent of $z$. This will be proven by finding
an analog of the Callan-Symanzik equation of field theory, that describes
the change of the two-point function of a composite
operators under change of the UV cutoff. This equation will show,
that the low-momentum limit of the two-point function is only sensitive to
the IR region of the 5-d geometry --i.e. the region close to the brane.
We will also show that
the high-momentum limit of the two-point function is sensitive only to
the UV region of the 5-d geometry.

Let us start by recalling that in the geometry dual of field theory the
two-point function of a composite operator, ${\cal O}(x^\mu)$,
with source $\psi(x^\mu)$ is found as follows\footnotemark.
\footnotetext{Here, for simplicity, we will restrict ourselves
to minimally-coupled scalar fields.}
\begin{enumerate}
\item The field $\psi$ is promoted to a 5-d field $\psi(x^\mu,z)$. It
obeys some boundary conditions, given in refs.~\cite{gkp,w1}. Here we find
it convenient to follow the prescription of ref.~\cite{gkp}, and to
choose as boundary conditions at small $z$ a plane 4-d wave: $\psi(x^\mu)=
\exp(ik_\mu x^\mu)$.
\beq
\psi(x^\mu,z)=e^{ik_\mu x^\mu}\psi_k(z),\;\;\;
\psi_k(z)|_{z=\epsilon}=1, \;\;\; \lim_{z\rightarrow \infty}
\psi_k(z)=0, \;\;\; k_\mu k^\mu\equiv k^2 >0.
\eeq{m4}

A few comments are in order here. a) Here $\epsilon$ is an UV regulator, and
must be chosen much smaller than any other length scale in the problem, in
particular, $k \epsilon\ll 1$. b) When $k^2<0$
the boundary condition at large $z$ is $\psi_k(z) \propto
\exp(i|k|z + ik_\mu x^\mu)$.
c) If the 5-d geometry is not $AdS_5$ at large $z$, but
rather it develops a singularity at finite $z=a$, then the boundary
condition at $a$ is that $\psi_k(z)$ is regular near $z=a$.
In this paper, we set aside the latter possibility, and assume
that the large-$z$ geometry of the 5-d space obeys Eq.~(\ref{m3}).

\item The 5-d field $\psi_k(z)$ obeys free scalar equation of motion:
\beq
\left[ -\partial_z \partial_z -3\phi_z(z)\partial_z + k^2
+e^{2\phi(z)}M^2(z)\right]
\psi_k(z)=0.
\eeq{m5}
Here, $\phi_z(z)\equiv \partial_z\phi(z)$. The square-mass term $M^2(z)$
becomes constant both in the IR and in the UV
\beq
\lim_{z\rightarrow 0}M^2(z)=M^2_{ {{UV}} },\;\;\; \lim_{z\rightarrow \infty}
M^2(z)=M^2_{{{IR}} }.
\eeq{m6}
Generically, $M^2_{{{UV}} }\neq M^2_{{{IR}}  }$.
\item Finally, the two-point function $A(k^2)=\int d^4x \exp(ik_\mu x^\mu)
\langle {\cal O}(x){\cal O}(0)\rangle$ is given by~\cite{gkp}
\beq
A(k^2)= \left[ e^{3\phi(z)}\psi_k^*(z)\partial_z
\psi_k(z)\right]^\infty_\epsilon\equiv  \left[ e^{3\phi(z)}
{\partial_z\psi_k(z)\over \psi_k(z)}
\right]^\infty_\epsilon.
\eeq{m7}
The latter form of $A(k^2)$ is independent of the normalization condition
at $z=\epsilon$, and it is valid whenever $\psi_k(z)$ obeys the
correct boundary condition at $z=\infty$.
\end{enumerate}

We want to prove, first of all,
that the low-momentum behavior of $A(k^2)$ is insensitive to the
small-$z$ region of the 5-d geometry.

The key to the proof is a first-order equation for the two-point function,
somewhat reminiscent of the Callan-Symanzik equation.
Using Eq.~(\ref{m5}), and defining
\beq
A(k^2,z) \,=\, -  e^{3\phi (z)}\partial_z\log \left[ \psi_k(z)\right] \, ,
\eeq{def}
one finds the equation:
\beq
\partial_z A(k^2,z) \,=\,e^{-3\phi (z)} A^2(k^2,z) \,-\,
 k^2 e^{3 \phi (z)} \,-\, M^2(z) e^{5\phi(z)}.
\eeq{m8}
Notice that the boundary conditions at large $z$ ensure that
$\lim_{z\rightarrow \infty}A(k^2,z)=0$.  The two-point function
 plays the role of the
initial condition for Eq.~(\ref{m8}):
\beq
A(k^2,\epsilon)=A(k^2).
\eeq{m9}

Notice that, analogously to the Callan-Symanzik equation, Eq.~(\ref{m8}) does
not fix the value of the two-point function: that comes from solving
the second-order equation Eq.~(\ref{m5}), subject to the boundary conditions
given in Eq.~(\ref{m4}). Eq.~(\ref{m8}) describes instead the evolution of
a quantity, $A(k^2,z)$ that coincides with the true two-point function at small
$z$.

It is tempting to interpret $A(k^2,z)$ as the two-point function computed with
a cutoff $z$. This is indeed true if for low momenta, $k\rightarrow 0$,
$A(k^2,z)$ differs from $A(k^2)$ by at most a multiplicative factor,
and an additive factor either polynomial in $k^2$ or of higher order
in the $k^2$ expansion:
\beq
A(k^2,z)= Z^2(z) A(k^2) + P(k^2,z) + O\left[ k^2z^2 A(k^2)\right], \;\;\; kz
\ll 1 .
\eeq{m10}
The multiplicative factor $Z^2 (z)$ is interpreted as the wave-function
renormalization of the operator ${\cal O}(x^\mu)$. The polynomial $P(k^2,z)$
changes only the contact terms in the two-point function, without affecting
its behavior at non-coincident points. The last term is negligible in the
infrared limit.

Notice that, whenever Eq.~(\ref{m10}) holds, the dependence of the two-point
function on the UV geometry, i.e. the small-$z$ region, is completely
factored into contact terms and a wave-function renormalization constant.
In field theory, the same can be said {\em verbatim} for the dependence of
the two-point function on the UV cutoff. Therefore, if Eq.~(\ref{m10}) holds,
the coordinate $z$ plays exactly the role of a length cutoff.
This is another manifestation of the UV/IR connection for
non-conformal
theories. More interestingly, Eq.~(\ref{m10}) says that
in geometries with two AdS regions,
as in the examples in refs.~\cite{gppz,dz}, the infrared behavior of the
4-d theory is completely described by the IR
$AdS_5$ geometry, given in Eqs.~(\ref{m1},\ref{m3}). To study the IR, one can
ignore the behavior of the metric in the UV region $kz>1$.
As a concrete application of this result, the quantity
$M^2_{{{IR}} }$ is related to the IR
scaling dimension of ${\cal O}$, $\Delta_{{{IR}}  }$, by the standard AdS
formula~\cite{gkp,w1,ffz}
\beq
\Delta_{{{IR}} } \,=\, 2 \,+\, \sqrt{4+M^2_{ {{IR}} }
R^2_{{{IR}} }}.
\eeq{m11}

Eq.~(\ref{m10}) is easily proven. It is sufficient to notice that $A(k^2,z)$
is a smooth function of the initial conditions $A(k^2)$. This is a standard
property of ordinary differerential equations with smooth coefficients, as
Eq.~(\ref{m10}). A proof of this theorem can be found in~\cite{a}.

We are interested in the $k$-dependence of $A(k^2,z)$. Smoothness in the
initial conditions (and $z$) implies $A(k^2,z)= F(A(k^2),z,k^2)$, with
$F(A,z,k^2)$ a
smooth function of $A$, $z$ and $k^2$. The field-theory interpretation of
$A(k^2)$
tells us that the small-$k^2$ expansion of $A(k^2)$ reads~\footnotemark
\footnotetext{Here we write the expression valid for generic, non-integer
$\Delta_{{{IR}} }$. For integer $\Delta_{{{IR}} }$, the non-analytic term reads
$k^{2(\Delta_{{{IR}} }-2)}\log k^2$.}
\beq
A(k^2)=A(0)+Q(k^2) + c k^{2\Delta_{{{IR}} }-4} +
 O(k^{ 2\Delta_{{{IR}}}-2}).
\eeq{m12}
Here $c$ is a nonzero constant, positive by unitarity of the 4-d IR theory;
$Q(k^2)$ is a polynomial in $k^2$, vanishing at $k^2=0$.

By expanding $A(k^2,z)$ near the initial condition $A(0)$ we find
\begin{eqnarray}
A(k^2,z) &=& F(A(0),z,k^2) + {\partial F \over \partial A
}(A,z,k^2)\arrowvert_{A=A(0)} [A-A(0)]
+O\{[A-A(0)]^2\}\nonumber \\
&=& F(A(0),z,0) + R(k^2,z)+{\partial F \over \partial A
}(A,z,0)\arrowvert_{A=A(0)} [A-A(0)]+\nonumber \\ &&
+O\left\{ \left[ A-A(0)\right]^2, [A-A(0)]^2k^2\right\},
\label{m13}
\end{eqnarray}
where $R(k^2,z)$ is a polynomial in $k^2$, vanishing at $k^2=0$.
Eq.~(\ref{m10}), with $Z^2=\partial F/\partial A \arrowvert_{A=A(0)}$,
follows immediately from the expansion in Eq.~(\ref{m12})
and the smoothness of $F(A,z,k^2)$.

Positivity of the wave-function renormalization $Z$ is proven as follows.
By analytic continuation in $k^2$, $A(k^2,z)$ becomes an analytic function
with a cut along the real negative axis. By its definition, given in
Eq.~(\ref{def}),
it obeys $A(k^{2\,*},z)=A^*(k^2,z)$. By splitting Eq.~(\ref{m8}) into
real and imaginary part, we find the equation
\beq
\partial_z \Im A(k^2,z)= 2 e^{-3\phi (z)} \Re A(k^2,z) \Im A(k^2,z).
\eeq{m14}
Expanding its solution near $k^2=0$ we find~\footnotemark,
thanks to Eq.~(\ref{m12}):
\footnotetext{Again, we write an equation valid for non-integer
$\Delta_{{{IR}} }$. It is trivial to see that the equation for $\Delta_{{{IR}}
}$
integer gives the same result for the $Z$ factor.}
\beq
\Im A(k^2,z)= e^{2\int_\epsilon^z e^{-3 \phi (w)}\Re A(0,w) dw}
c\sin (4\Delta_{IR}\pi)
k^{2\Delta_{IR}-4} + O(k^{2\Delta_{IR}-2}).
\eeq{m15}
{}From this equation, it follows that the wave-function renormalization factor
is positive:
\beq
Z(z) \,=\, e^{\int_{\epsilon}^z e^{-3 \phi (w)}\Re A(0,w) dw}.
\eeq{m16}

After having studied the IR limit of the two-point function, we want to
study the opposite limit, namely $k^2\rightarrow \infty$. We want to
show that, in this limit, $A(k^2)$ can be computed by approximating the
metric~(\ref{m1}) with its UV AdS form, given in Eq.~(\ref{m2}).
We shall do so by writing Eq.~(\ref{m5}) in the form of an integral equation
and studying its ``Jost solution'' in the limit $k^2\rightarrow \infty$.

Let us write $\phi (z)$ in the form
\beq
\phi (z) \,=\, - \log \left( \frac{z}{R_{{{UV}}}}\right)+h(z),
\eeq{a1}
where $h(z)$ has properties
\beq
h(z) =O(z) \, \, \; \;  \mbox{for} \, z\rightarrow 0 ,  \; \; \;
h(z)  = \log \left(\frac{R_{{{IR}}}}{R_{ {{UV}} }}\right) + O(1/z) \; \; \;
\mbox{for} \, z\rightarrow \infty  .
\eeq{a2}
By writing $\psi_k = e^{-\frac{1}{2}\phi (z)}f_k(z)$,  Eq.~(\ref{m5})  is
converted into the Schr\"{o}dinger equation
\beq
f_k'' -\frac{15}{4 z^2} f_k - k^2 f_k -
\frac{M^2_{ {{UV}}  }R^2_{{{UV}}   }}{z^2} f_k = V(z) f_k,
\eeq{a3}
where
\beq
V(z) = \frac{3}{2}h'' +  \frac{9}{4}(h')^2 - \frac{9}{2 z} h' -
\frac{M^2_{{{UV}}   }R^2_{{{UV}}   }}{z^2}\left(1 - \frac{M^2(z)}{M^2_{ {{UV}}
}}e^{2h(z)}\right).
\eeq{a4}
Here
\beq
V(z)  =  O(1/z) \; \; \; \, \mbox{for} \, z\rightarrow 0  , \; \; \; \;
V(z) =  \frac{\delta}{z^2} +   O(1/z^3) \; \; \;  \mbox{for} \, z
\rightarrow \infty  ,
\eeq{a5}
where $\delta = M^2_{{{IR}}}R^2_{{{IR}}}-M^2_{{{UV}} }R^2_{{{UV}}}$.
The pure-AdS Eq.~(\ref{a3})  --  with $V(z)=0$  --  
has two independent solutions :
$\sqrt{2 \pi k z}K_{\nu}(k z)$ and  $\sqrt{2 \pi k z}I_{\nu}(k z)$. We are
looking for the solution of the full equation defined by the
 boundary condition $\lim_{z\rightarrow \infty}e^{kz}f_k(z) =1$.
Constructing an appropriate Green's function, we can write Eq.~(\ref{a3})
(or Eq.~(\ref{m5})) in the form of an integral equation. Let
$\psi_k(z)=e^{-\frac{3}{2}h(z)}\tilde{\psi_k}(z)$, then $\tilde{\psi_k}(z)$ is
a solution of
\beq
\tilde{\psi_k}(z) \,=\, \frac{k^2z^2}{2}K_{\nu}(kz) \,-\, z^2
\int_{z}^{\infty}\frac{d\xi }{\xi}G(\xi , z; k) V(\xi )\tilde{\psi_k}(\xi ),
\eeq{a6}
where
\beq
G(\xi , z; k) \,=\, I_{\nu}(kz)K_{\nu }(k\xi )-  I_{\nu }(k\xi )K_{\nu }(kz).
\eeq{a7}
We normalize the solution $\tilde{\psi_k}(z)$ to 1 at $z=\epsilon$:
\beq
\Psi_k (z, \epsilon ) \,=\, \frac{\psi_k (z)}{\psi_k (\epsilon )},
\eeq{a8}
then the two-point function
is given by
\beq
A(k^2) \,=\, \left[ e^{3\phi(z)}
{\partial_z\Psi_k(z, \epsilon )\over \Psi_k(z, \epsilon )}
\right]^\infty_\epsilon  \,=\, \left[ e^{3\phi(z)}
{\partial_z \tilde{\psi_k}(z)\over \tilde{\psi_k}(z)} - e^{3\phi(z)} {3\over
2}h'(z)\right]^\infty_\epsilon .
\eeq{a9}
All the $k$-dependence in Eq.~(\ref{a9}) is contained in
$\partial_z\log[\tilde\psi_k]$.
It is therefore sufficient to compute $\partial_z \tilde{\psi_k}(z) /
\tilde{\psi_k}(z)$. The solution of  Eq.~(\ref{a6}) is given by the series
\beq
\tilde{\psi_k}(z) \,=\, \tilde{\psi_k}^{(0)}(z) \,+\,  \tilde{\psi_k}^{(1)}(z)
\,+\, \dots ,
\eeq{a10}
where
\beq
\tilde{\psi_k}^{(0)}(z) \,=\, {k^2z^2\over 2}K_{\nu} (kz),
\eeq{a11}
\beq
\tilde{\psi_k}^{(n+1)}(z) \,=\, - {k^2z^2\over 2}\int_{z}^{\infty}\frac{d\xi
}{\xi}G(\xi , z; k) V(\xi )\tilde{\psi_k}^{(n)}(\xi ).
\eeq{a11a}
It can also be written as
\beq
\tilde{\psi_k}(z) \,=\, a(z,k) \; {k^2z^2\over 2}K_{\nu} (kz) \,+\, b(z,k) \;
{k^2z^2\over 2}I_{\nu} (kz),
\eeq{a12}
where
\beq
 a(z,k) \,=\, 1 \,+\, {2\over k^2}\int_{z}^{\infty}\frac{d\xi }{\xi} V(\xi
)I_{\nu}( k\xi )\tilde{\psi_k}(\xi ),
\eeq{a13}
\beq
 b(z,k) \,=\,- {2\over k^2}\int_{z}^{\infty}\frac{d\xi }{\xi} V(\xi )K_{\nu}(
k\xi )\tilde{\psi_k}(\xi ).
\eeq{a14}
Using  Eq.~(\ref{a10}) we can write $a(z,k)=1+a_1+\dots \, $,  $ \;
b(z,k)=b_1+\dots$,
where
\beq
a_1 \,=\, {1\over k} \int_{\mu}^{\infty} {t\over k} V\left( {t\over k}\right)
I_{\nu}(t)K_{\nu}(t)dt,
\eeq{a15}
\beq
b_1 \,=\,- {1\over k} \int_{\mu}^{\infty} {t\over k} V\left( {t\over k}\right)
K_{\nu}^2(t)dt,
\eeq{a16}
etc., and $\mu = kz$. Here, we are interested in the $\mu \ll 1$ case, since
we want to study the region $kz \sim \epsilon \ll 1$.

The integrals are dominated by the contribution of the $t \sim \mu \ll 1$
region where we have $zV(z)= O(1)$ and, therefore,
\beq
a_1 \,=\, \frac{{\cal A}(\mu )}{k} \,+\, O\left( 1/k^2 \right),
\eeq{a17}
\beq
b_1 \,=\, - \frac{{\cal B}(\mu )}{k} \,+\, O\left( 1/k^2 \right).
\eeq{a18}
In general, we have $a_n\sim O(1/k^n)$, $b_n\sim O(1/k^n)$. We arrive,
therefore, at the standard Born-type series for $a$ and $b$.

Let us see now how the two-point function depends on $a$ and $b$. We have
\beq
{\partial_z \tilde{\psi_k} \over \tilde{\psi_k}}\,=\, {\nu +2\over z} \,-\,
{k (K_{\nu +1}(kz) - r(k,z)I_{\nu +1}(kz))\over K_{\nu}(kz) +
 r(k,z)I_{\nu}(kz)} \,+\, { a'\over a} { K_{\nu}(kz) + {b'\over a'}
I_{\nu}(kz)\over K_{\nu}(kz) +
 r(k,z)I_{\nu}(kz)},
\eeq{a19}
where $r(k,z)=b(k,z)/a(k,z)$ and $a'=\partial_z a$, $b'=\partial_z b$.
Let us consider massive and massless cases separately.
\subsubsection*{Massive case, $\nu > 2$}
We have
\beq
K_{\nu}(kz) +
 r(k,z)I_{\nu}(kz) \,=\, {2^{\nu - 1}\Gamma (\nu) \over  (kz)^{\nu}}\left( 1
+\cdots - c_{\nu}(k,z) \left({kz\over 2}\right)^{2 \nu}{\Gamma (1-\nu )\over
\Gamma (1+\nu )} +\cdots \right),
\eeq{a20}
where
\beq
c_{\nu}(k,z)\,=\, 1\,-\, {2 r(k,z)\over \Gamma (\nu )\Gamma (1-\nu )},
\eeq{a21}
and
\beq
K_{\nu +1}(kz) -
 r(k,z)I_{\nu +1}(kz) = {2^{\nu}\Gamma (\nu +1) \over
 (kz)^{\nu +1}}\left( 1 +\cdots - c_{\nu +1}(k,z)
 \left({kz\over 2}\right)^{2 \nu +2}{\Gamma (-\nu )\over \Gamma (\nu +2 )}
+\cdots \right),
\eeq{a22}
so
\beq
{\partial_z \tilde{\psi_k} \over \tilde{\psi_k}}\,=\,- {2\Gamma (1-\nu )
\over z \Gamma (\nu )}\left({kz\over 2}\right)^{2 \nu} \,+\,
 {4 r(k,z)\over z \Gamma^2 (\nu )}\left[ 1 + {z\over 2 \nu }{r'(k,z)\over
r(k,z)}\right]\left({kz\over 2}\right)^{2 \nu} \,+\, \cdots ,
\eeq{a24}
where $r'(k,z)=\partial_z r(k,z)$ and $r(k,z)=b/a = O(1/k)$, and dots represent
higher powers of $k$.
The $k$-dependent part of $A(k^2)$ is
\begin{eqnarray}
A(k^2)  &=& R_{{{UV}}  }^3 e^{3 h(\epsilon )}
 {2\Gamma (1-\nu )\over  \epsilon^4 \Gamma (\nu )}\left({k\epsilon \over
2}\right)^{2 \nu}\left( 1 \,+\, {2 r\over \Gamma (\nu ) \Gamma (1-\nu )}\left[
1\,+\, {\epsilon\over 2 \nu}{r'\over r}\right] + O\left( 1/k^2 \right) \right)
\nonumber \\ &+& O\left( (k\epsilon )^{2\nu +n} \right),
\label{a25}
\end{eqnarray}
where $n>0$. The term proportional to $r(k,z)$ is the correction to the pure
AdS result obtained in~\cite{gkp}.

\subsubsection*{Massless case, $\nu = 2$.}

Here
\beq
{\partial_z \tilde{\psi_k} \over \tilde{\psi_k}}\,=\, -{k(K_3 (kz) - r I_3(kz)
)\over K_2 (kz) + r I_2(kz)} \,+\, {a'\over a}{K_2 (kz) + {b'\over a'} I_2(kz)
\over K_2 (kz) + r I_2(kz)}.
\eeq{a26}
For $a'=0$, $a=1$, $r=0$ this gives~\cite{gkp}
\beq
{\partial_z \tilde{\psi_k} \over \tilde{\psi_k}}\,=\, -{k K_3 (kz)\over K_2
(kz)} \,=\, -{z^3 k^4 \over 4} \log k \,+\, O\left( k^6z^5 \right).
\eeq{a27}
For nonzero $r$,
\beq
{\partial_z \tilde{\psi_k} \over \tilde{\psi_k}}\,=\, -{z^3 k^4 \over 4} \log k
\left( 1 \,+\, {a'z\over 4a} \,+\, O\left( k^2z^2, 1/k^2 \right)\right).
\eeq{a28}
Here $a'/a = O(1/k)$.
Therefore,
\beq
A(k^2) \,=\, R^3_{{{UV}}  }e^{3h(\epsilon )}{k^4\over 4}\log k \left( 1\,+\,
{a'\epsilon \over 4 a}\,+\, \cdots \right),
\eeq{a29}
to leading order in $k\epsilon$ and $1/k$.

We can see, therefore, that in the $k^2\rightarrow \infty$ limit the two-point
function can be written as
\beq
A(k^2) \,=\, \tilde{Z}^2 (\epsilon )A_{AdS}(k^2)\left( 1 \,+\, O(1/k) \right),
\eeq{a30}
where $\tilde{Z}^2 (\epsilon )=e^{3h(\epsilon )}$
depends on interpolating metric, but not on $k$.

In summary, in this paper we have shown how Green's functions of
composite operators in geometry duals of strongly coupled gauge theories can
be computed in the IR and UV limit. We have found the pleasant result that
the IR behavior of the Green's functions depends only on the ``near-brane''
geometry. We have also checked that, in all theories that are UV asymptotic
to N=4, 4-d super Yang-Mills, the UV Green's functions are universal.

It would be interesting to see if this result can be strenghtened to include
the case when string corrections to classical gravity become important.
It would be also important to see whether the formal similarity between
Eq.~(\ref{m8}) and the Callan-Symanzik equation is an accident, or it suggests
instead a way of writing the C-S equation in geometry duals of gauge theories.
\vskip .2in
\noindent
{\bf Acknowledgements}\vskip .1in
\noindent
M.P. is supported in part by NSF grant no. PHY-9722083.

\end{document}